\documentclass[a4paper,11pt]{article}
\usepackage{pos}

\title{Multipolynomial Monte Carlo Trace Estimation}

\author[a]{Paul Lashomb}
\emailAdd{paul\_lashomb@baylor.edu}
\author[a]{Ronald B. Morgan}
\emailAdd{ronald\_morgan@baylor.edu}
\author[c]{Travis Whyte}
\emailAdd{whytetr@tcd.ie}
\author*[a]{Walter Wilcox}
\emailAdd{walter\_wilcox@baylor.edu}

\affiliation[a]{Department of Physics, Baylor University, Waco, TX 76798-7316, USA}

\affiliation[b]{School of Mathematics, Trinity College, College Green, Dublin 2, IRE}

\abstract{In lattice QCD the calculation of disconnected quark loops from the trace of the inverse quark matrix
has large noise variance. A multilevel Monte Carlo method is proposed
for this problem that uses different degree polynomials on a multilevel system. The
polynomials are developed from the GMRES algorithm for solving linear
equations. To reduce orthogonalization expense, the highest degree polynomial
is a composite or double polynomial found with a polynomial preconditioned
GMRES iteration. Matrix deflation is used in three different ways: in the Monte Carlo levels, in the main solves, and in the deflation of the highest level double polynomial. A numerical comparison with optimized Hutchinson is performed on a quenched $24^4$ lattice. The results demonstrate that the new Multipolynomial Monte Carlo method can significantly improve the trace computation for matrices that have a
difficult spectrum due to small eigenvalues.}

\FullConference{The 40th International Symposium on Lattice Field Theory (Lattice 2023)\\
July 31st - August 4th, 2023\\
Fermi National Accelerator Laboratory\\}

\begin{document}
\maketitle

\section{Introduction}

Disconnected quark loop operators on the lattice come in form of a trace,
\begin{equation}
    \langle \bar{\psi} \Theta \psi \rangle = -Tr(\Theta M^{-1}).
\end{equation}\noindent
This is a challenging problem that is generally solved with Monte Carlo or noise sampling. We give a new method that uses multilevel Monte Carlo with different degree polynomials used to form the levels. It also incorporates eigenvalue deflation in various crucial ways. We show that this approach can significantly improve computation compared to simple Monte Carlo in which $Tr(\Theta M^{-1})$ is approximated with a Monte Carlo method using $N$ Hutchinson samples: $\frac{1}{N}\sum _{i=1}^{N} \eta^{\dagger}_i\Theta M^{-1}\eta_i$~\cite{HutchTrace}. We will use $\eta_i$ as a vector of $Z(4)$ noise, but other choices are possible.  A large system of linear equations $Mx_i = \eta_i$ must be solved for each sampling, and this is the main expense of the method. 

\section{Subtraction Methods}

Subtraction methods are needed in order to reduce the variance of noisy calculations. The noise subtracted trace estimator is
\begin{equation} \label{eq:trace_simple}
    \begin{aligned}
    Tr(\Theta M^{-1}) = & \frac{1}{N}\sum_{n=1}^N
    \eta^{(n)\dagger}\Theta \left(  M^{-1} - \tilde{M}^{-1}\right)\eta^{(n)} + Tr(\Theta\tilde{M}^{-1}),
    \end{aligned}
\end{equation}\noindent
where $M^{-1}$ is the inverse quark matrix and $ \tilde{M}^{-1}$ is the subtracted matrix which reduces the variance\cite{Thron}. The second term in Equation~(\ref{eq:trace_simple}) is a trace correction term which does not affect the variance. This forms the basis for all of our subtraction methods, each of which involve a different approximation, $\tilde{M}^{-1}$, for the inverse of the Wilson-Dirac matrix $M^{-1}$. 

\section{The GMRES Polynomial}

We use a polynomial, $\pi(M)$, formed from the GMRES algorithm\cite{morgan19}. It is given by
\begin{equation}
    \pi(M) = \prod_{i = 1}^{d} \Big( 1 - \frac{M}{\tilde{\theta}_i} \Big), 
\end{equation}
where $\tilde{\theta}_1, \tilde{\theta}_2, \ldots, \tilde{\theta}_d$ are the Leja ordered \cite{reichel}, harmonic Ritz values obtained from a single cycle of GMRES($d$), and $d$ is the desired degree of the GMRES polynomial. This method of finding the polynomial is much more stable than using the normal equations\cite{QuanLiu}, allowing for high degree polynomials to be formed. 

The GMRES polynomial $p(M)$ of degree $d-1$ related to $\pi(M)$ is given by $\pi(M)=1-M\,p(M)$. A backwards solve shows it can be expressed as
\begin{equation}
p(M)=\sum_{k=1}^{d}u_k,\,\, u_k=\frac{M}{\tilde{\theta}_k}  \left( 1-  \frac{M}{\tilde{\theta}_1} \right)  \left( 1-  \frac{M}{\tilde{\theta}_2} \right) \cdots  \left( 1-  \frac{M}{\tilde{\theta}_{k-1}} \right),
\end{equation}
where $d$ is the order of the polynomial $\pi(M)$. We only need to run a single cycle of GMRES($d$) in order to extract $d$ harmonic Ritz values. These are then Leja ordered for numerical stability. For subtraction, we make the assignment $\tilde{M}^{-1}\equiv p(M)$.

\section{Relative Variance}

In order to monitor the reduction in variance of each method over that of no subtraction (NS), we define the relative variance as
\begin{equation}
    \bar{\sigma}_{R}^2 \equiv \frac{\bar{\sigma}_{A}^2}{\bar{\sigma}_{NS}^2},
    \label{eq:relvar}
\end{equation}
where $\bar{\sigma}_A^2$ is the variance obtained from a particular subtraction method $A$ averaged over the configurations, and $\bar{\sigma}^2_{NS}$ is the variance of the trace estimator with no subtraction averaged over all configurations.

The standard relative variance results are often dominated by a single configuration and have large fluctuations. Log-averaged relative variance results, although not unbiased, have significantly reduced errors and are far easier for other numerical groups to make comparisons to. Thus, we also define a base-10 log-averaged relative variance using the geometric mean \cite{sheskin04} as
\begin{equation}
    \bar{\sigma}_{R,log}^2 \equiv 10^{\bar{\rho}^2_{A} - \bar{\rho}^2_{NS}},
    \label{eq:logaverelvar}
\end{equation}
where $\bar{\rho}^2_{A} = \frac{1}{N}\sum_{j=1}^{N} \log_{10}[(\sigma_{A}^2)^{(j)}]$ ($\bar{\rho}^2_{NS}$ defined similarly), $(\sigma_{A}^2)^{(j)}$ is the variance obtained for configuration $j$ for a particular method, and $N$ is the number of configurations.

We now turn our attention to considering the polynomial degree dependence of our polynomial noise subtraction methods. We first examined both the standard and log-averaged relative variance for several polynomial degrees on different lattice volumes\cite{PolyQCD}. Here we consider the scalar operator but with four quenched lattices of volumes $4^3 \times 4$, $8^3 \times 8$, $12^3 \times 16$ and $24^3\times 32$ with $\beta = 6.0$ configurations and hopping parameter $\kappa = 0.1570 \approx \kappa_{crit}$ using a total of 10 thermalized configurations for each lattice volume. The three smaller lattices used $N = 10$ unpartitioned $Z(4)$ noise vectors for the Monte Carlo while the largest used $N = 100$. $N=100$ was sufficient to reveal the relative variance behavior for the $24^3\times 32$ lattices at approximately at the same relative variance error level as the $12^3\times 16$ lattices.

\begin{figure}[h!]
    \centering
    \includegraphics[scale=0.7]{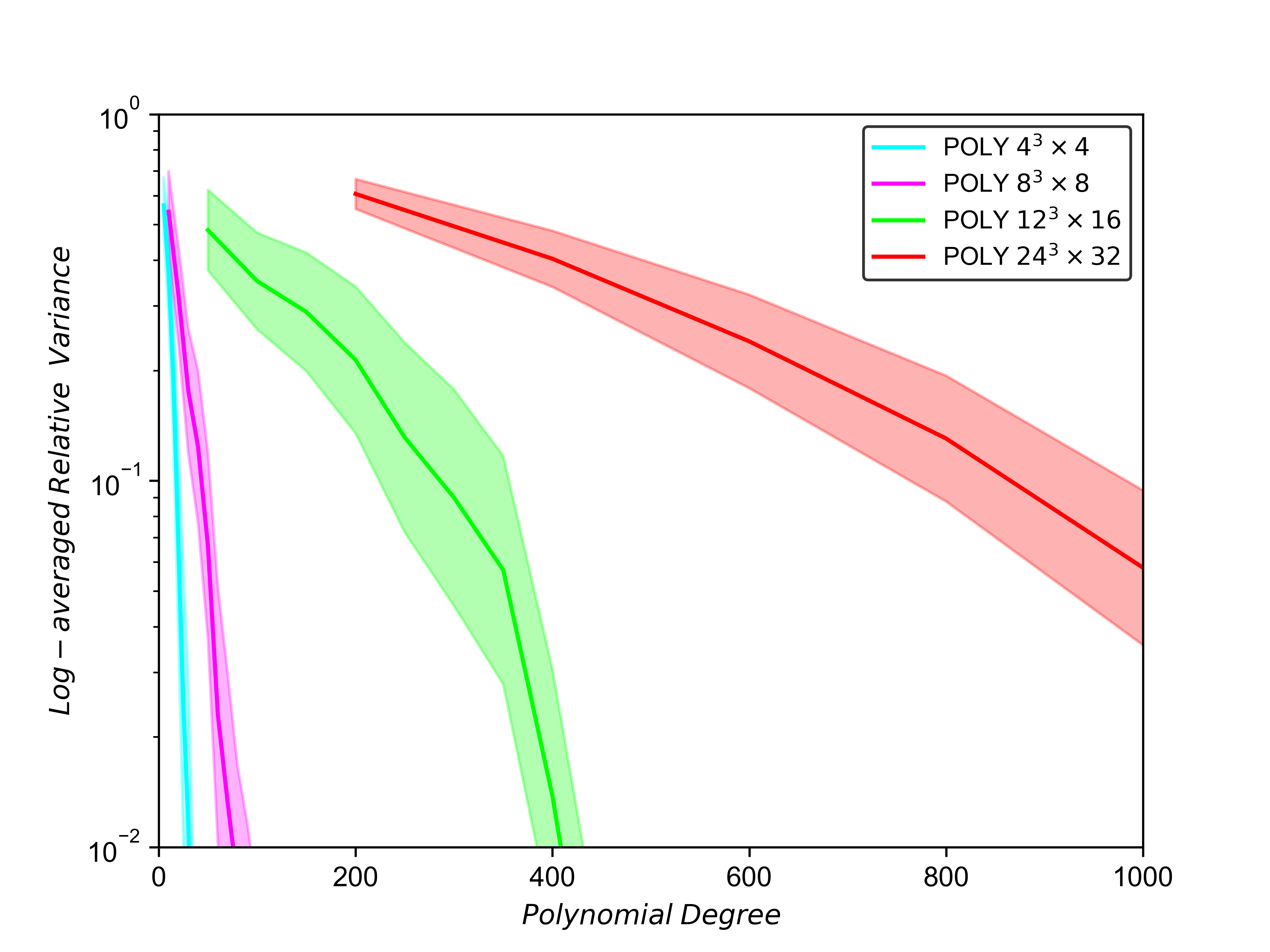}
    \caption{Log-averaged standard relative variance of the scalar operator using polynomial subtraction against the degree of the subtraction polynomial. Lattice volumes $4^3 \times 4$, $8^3 \times 8$, $12^3 \times 32$, and $24^3 \times 32$ are shown, each averaged over 10 quenched configurations at $\beta = 6.0$ and $\kappa = 0.1570$.}
    \label{fig:re_var_deg_allvols}
\end{figure}

Figure~\ref{fig:re_var_deg_allvols} shows the log-averaged standard relative variance of the scalar operator versus the polynomial degree for each of the four lattice volumes that we used.  Error bars are indicated by the shaded regions. Note the approximate exponential falloff in the relative variances for larger polynomial degrees for either figure. We also see that the increase in lattice volume requires a higher polynomial degree to achieve the same variance reduction performance. This polynomial degree can become impractically high for larger lattices. To increase the polynomial degree further, one would need to reduce the expense of computing such high degree polynomials as the single cycle of GMRES($d+1$) involved in forming the subtraction polynomial of degree $d$ becomes expensive due to the orthogonalization costs of the Arnoldi iteration. As will be seen shortly, double polynomials as discussed in Refs.~\cite{embree21} and \cite{morgan19}, and described in the next section can be used to combat this.

\section{Reducing Costs Through Double Polynomials}
The degree of the subtraction polynomials are dependent on the size of the Krylov subspace used to form them. Since the GMRES algorithm is based on the Arnoldi iteration, forming large subspaces to produce high degree subtraction polynomials can increase the orthogonalization and memory costs considerably. As a cost-saving measure, we may make use of double polynomials to avoid the large orthogonalization and memory costs of the Arnoldi iteration. We start this section by first outlining double polynomials in the context of polynomial preconditioning as they were originally \cite{embree21} used, and then we state how they can be used in noise subtraction. 

In polynomial preconditioning, we use a polynomial preconditioner $p_{in}(M)$ on the system $M$, yielding the polynomial preconditioned system $\phi_{in}(M) \equiv M p_{in}(M)$ as
\begin{equation}
     \phi_{in}(M) y = b,  
\end{equation}
where the solution is $x = p_{in}(M) y$. 
If we then polynomial precondition the system a second time using the preconditioner $p_{out}(\phi_{in}(M))$, we will have a double polynomial preconditioned system
\begin{equation}
     \phi_{out}(\phi_{in}(M)) z \equiv \phi_{in}(M) p_{out}(\phi_{in}(M))z = b,  
\end{equation}
where the solution is now $x = p_{in}(M) p_{out}(\phi_{in}(M))z$. If we now consider 
\begin{equation}
   \phi_{out}(\phi_{in}(M)) = M p_{in}(M) p_{out}(\phi_{in}(M)),  
\end{equation}
we see that performing double polynomial preconditioning can also be thought of as applying a single polynomial preconditioner $p_{double}(M) \equiv p_{in}(M) p_{out}(\phi_{in}(M))$ to the original system $M$. 

To form such a double polynomial, we perform a single cycle of GMRES($d_{in}+1$) on $M$ to obtain $d_{in}+1$ Leja ordered, harmonic Ritz values for $M$, and then perform a single cycle of GMRES($d_{out}+1$) on $\phi_{in}(M)$ to obtain $d_{out}+1$ Leja ordered, harmonic Ritz values for $\phi_{in}(M)$. Note that the latter single cycle of GMRES($d_{out} + 1$) on $\phi_{in}(M)$ can alternatively be thought of as performing a single cycle of polynomial preconditioned GMRES or PP-GMRES($d_{out}+1$) on $M$ to form the $d_{out}+1$ Leja ordered, harmonic Ritz values. From these, one can form polynomials $p_{in}(M)$ and $p_{out}(\phi_{in}(M))$ of degrees $d_{in}$ and $d_{out}$, respectively, using the algorithm discussed in Ref.~\cite{embree21} to obtain a double polynomial of degree $d_{double} = (d_{in} + 1)*(d_{out} + 1) - 1$. 

In this way, we may form two low-degree polynomials $p_{in}(M)$ and $p_{out}(\phi_{in}(M))$ using smaller Krylov subspaces to achieve an effective high degree polynomial, thus avoiding the large Krylov subspaces formed by the Arnoldi iteration of GMRES applied solely on $M$. In the context of the multipolynomial method described in the next section, we make use of this double polynomial for subtraction purposes, $\tilde{M}^{-1}_{poly} \equiv p_{double}(M)$, at the highest level in our multilevel Monte Carlo.

\section{Multipolynomial Monte Carlo}
Multilevel Monte Carlo~\cite{Gi08,Gi15} uses different levels of approximation to the problem being solved in order to reduce the variance.  It creates several independent Monte Carlo estimations with each except the last one having differences between levels.  The last has a Monte Carlo for the least accurate approximation to the original problem.  Essentially, much of the Monte Carlo sampling is shifted from the original problem to sampling with the approximations that are cheaper to use. Multilevel Monte Carlo has been applied in QCD~\cite{GiHaNaSc,WhStRoOr}.  Also, Ref.~\cite{FrKhRH} has levels with different size multigrid matrices. 

In~\cite{PolyQCD} it is suggested to use a high degree polynomial of $M$ to approximate $M^{-1}$, and in Ref.~\cite{MPolyQCD} to turn this into a multilevel Monte Carlo method. Consider
\begin{equation}
    Tr(M^{-1}) = Tr(M^{-1} - p_1(M) ) + Tr(p_1(M)). \label{onepolytr}
\end{equation}
The two traces on the right can be computed independently with Monte Carlo sampling.  Or if $p_1$ is low degree, the second trace can be computed exactly with probing~\cite{LaStOr,LaSt}.
If $p_1$ is a good approximation to $M^{-1}$, the variance is greatly reduced for the $Tr(M^{-1} - p_1(M))$ part of the Monte Carlo. Thus very few samples are needed for this part of the trace estimate. Each sample is expensive however, needing solution of linear equations and application of a high degree polynomial. We use the stable high degree polynomials that are generated by the GMRES algorithm in~\cite{embree21,morgan19}. However, Equation \eqref{onepolytr} cannot easily be used with a high degree polynomial because finding $Tr(p_1(M))$ would require a Monte Carlo approach that has variance similar to that of the original problem of $Tr(M^{-1})$.  It does gain in that implementing $p_1(M)$ times a vector is generally less expensive than solving the linear equations needed to multiply $M^{-1}$ times a vector.  

To reduce the Monte Carlo cost for $Tr(p_1(M))$ we suggest approximating $p_1$  with a lower degree polynomial, say $p_2$. 
This is done by making $p_2(M)$ another approximation to $M^{-1}$ and thus also an approximation to $p_1(M)$.  The next equation shows how $p_2$ is used so that there are two Monte Carlo's along with the problem of finding the trace of $p_2(M)$:
\[
    Tr(M^{-1}) = Tr(M^{-1} - p_1(M) ) + Tr( p_1(M) - p_2(M) ) + Tr(p_2(M)). 
\]
This is a polynomial version of multilevel Monte Carlo~\cite{He98,Gi08,Gi15}. In~\cite{HaTr}, multiple Chebyshev polynomials are used in a multilevel Monte Carlo method, however only for symmetric matrices and not for computing the trace of an inverse.  As mentioned above, we instead use GMRES polynomials for our non-Hermitian complex QCD matrices. Our approach is perhaps easier to set up than a multilevel Monte Carlo method that uses multigrid matrices~\cite{FrKhRH} since it does not need the development of the matrices. However, there are also complications for our new approach such as the need of a high degree polynomial $p_1$ and the need for eigenvalue deflation.  

For each lower order polynomial $p_i$ we want $ p_i(M) \approx M^{-1}$. These polynomials can be generated from a run of GMRES. For example, for a degree 100 polynomial $p(\alpha)$, stop GMRES at iteration 101. We call the GMRES residual polynomial $\pi(\alpha)$.  It is degree 101. This polynomial in $\alpha$ can be written as $\pi(\alpha) = 1 - \alpha p(\alpha)$, where $p(\alpha)$ is the polynomial that approximates the inverse of $M$.  In fact, $\pi(0)=1$ and $\pi(\alpha) \approx 0$ over the spectrum of $M$, once GMRES has run far enough.  So $p(\alpha) \approx 1/\alpha$ over the spectrum and thus $p(M) \approx M^{-1}$.  

There can be any number of levels in a multilevel Monte Carlo scheme, however there is cost for extra levels because then each individual Monte Carlo sampling must be estimated more accurately.  We will only go as high as using three polynomials which gives four trace calculations.  
Let the polynomials $p_1$, $p_2$ and $p_3$ have degrees be $d_1 > d_2 > d_3.$  The multilevel Monte Carlo with polynomials for the different levels, which we refer to as Multipolynomial Monte Carlo, uses this formula:
\begin{multline}
    Tr(M^{-1}) = Tr(M^{-1} - p_1(M)) + Tr(p_1(M) - p_2(M) ) + \\ Tr(p_2(M) - p_3(M) ) + Tr(p_3(M) ). \label{eq:multpolytr}
\end{multline}
The first three trace computations on the right of Equation \eqref{eq:multpolytr} can be performed with Hutchinson Monte Carlo trace estimation. Here we do the last one exactly with probing.

\section{Deflation for the Monte Carlo}

Deflation of eigenvalues is used in three different ways in our method. There is deflation in the Monte Carlos levels that have differences between polynomials, deflation in the solution of linear equations, and in the development of a deflated double polynomial. Here we discuss the Monte Carlo level deflation. The more complete discussion is in Ref.~\cite{MPolyQCD}.

We use polynomial preconditioned Arnoldi method (PP-Arnoldi)~\cite{embree21} to deflate eigenvalues. In some of the tests, we determine the stopping point by monitoring the relative residual of the corresponding polynomial preconditioned GMRES. Multipolynomial Monte Carlo can reduce expense due to moving much of the noise sampling to the cheaper Monte Carlo parts such as $Tr(p_2(M) - p_3(M) )$. However, the improvement due to having more levels may not be dramatic due to the increased number of samples needed for more Monte Carlos with more demanding error tolerances. Needed for a big improvement is deflation in the Monte Carlos that have the difference between polynomials. We find eigenvalues $\lambda_i$ and corresponding right and left eigenvectors $z_i$ and $u_i$, respectively. Then the Monte Carlos for $Tr(p_1(M) - p_2(M) )$ and $Tr(p_2(M) - p_3(M) )$ have deflation applied. For the first of these, this is
\[Tr \Bigl(p_1(M) - p_2(M) - \sum_{i}(p_1(\lambda_i) - p_2(\lambda_i)) z_i u_i^*\Bigr).
\]  
The deflation reduces the variance for Monte Carlo sampling. We correct for the deflated part by finding its trace exactly using that the trace of $z_i u_i^*$ is the inner product between the vectors.  

Note that when simply doing a Monte Carlo for $Tr(M^{-1})$, deflating eigenvalues is not very effective because of non-normality effects. For a non-normal matrix, there can even be an increase in the norm of the inverse as some spectral components are removed. The norm of $M^{-1}$ will not increase if singular value decomposition components are instead removed, hence they have been previously used~\cite{QCDsubtr,QCDsubtr2,GaStOr,QCDsubtr3,RoStOr}. Nevertheless, we remove eigenvalue components, because the singular values and vectors of a polynomial of $M$ are not necessarily related to those of $M$. Also, eigenvalues work well in our context because the polynomials all approximate $M^{-1}$ but differ mostly at the small eigenvalues that we are deflating. So removing these components significantly reduces the variance.

\section{Algorithm and Testing}

We test with 10 quenched matrices from $24^4$ configurations using the scalar operator. The desired error is set at $\epsilon = 0.0005 * 24^4$. We use two versions of the multpolynomial approach. Both are with three polynomials.  The first polynomial is from deflated PP(50)-GMRES solved to residual norm below $10^{-5}$.  This gives $p_1(M)$ that is a very accurate approximation to $M^{-1}$, so the Monte Carlo for $Tr(M^{-1} - p_1(M))$ requires only two noises and achieves far more than the requested accuracy.  The second and third polynomials are degree 200 and degree 4. The standard error for the second Monte Carlo is checked after three noises and then rechecked after every three. The third Monte Carlo is checked after six noises and then every third thereafter.  
The first test computes the eigenvectors by running PP(50)-Arnoldi until the corresponding GMRES relative residual norm goes below $10^{-12}$.  From the subspace thus generated, 30 right eigenvectors are computed corresponding to the smallest Ritz vectors, the left eigenvectors are found and the eigenvalue is accepted if residual norms are below $10^{-3}$. Ritz values are identified as real if the imaginary parts are below $2*10^{-4}$. An average of 22.4 small eigenvalues are used. The time and matrix-vector products for computing $Tr(p_3(M))$ with probing are included in the table (the time averages 0.45 hours).
The time for the whole multipolynomial process varies greatly from 3.9 hours up to 16 hours per configuration. This is because the Monte Carlo deflation is sometimes more effective than for other cases.  The number of noise vectors needed for the second Monte Carlo varies from 3 to 66, and every noise vector is expensive because it requires multiplication by the high-degree polynomial $p_1$ (and by $p_2$).

Next, we check whether it is worth spending more effort developing the eigenvalues for deflation. We use a degree 69 polynomial for $p_{in}$, and run PP(70)-Arnoldi for 150 iterations (stopping at that set point instead of at a GMRES residual level).  This takes a total of 10{,}500 matrix-vector products which compares to an average of 6250 matrix-vector products used just above when going to $10^{-12}$. (This varies considerably for the 10 matrices, between 4850 and 7000).  Next, the 60 smallest Ritz values and vectors are tested for accuracy.  An average of 52.9 approximate eigenvalues are accepted.  Most of them are very accurate with residual norms around $10^{-15}$. This makes the decisions more clear-cut for determining which eigenvalues to use for deflation.  We call the procedure ``extra deflation". The overall time goes down for all tests even though more time is spent on the initial steps of finding eigenvalues and polynomials before the actual Monte Carlos.

We show a comparison of the two Multipolynomial Monte Carlo tests along with Hutchinson in Table 1 with averaging over 10 configurations. Linear equations are solved for Hutchinson using PP(70)-GMRES with extra deflation. The time is reduced by a factor of more than six with the extra deflation version of the multipolynomial method. We note that the Hutchinson method takes an average of 19.6 hours, but this would take much longer if polynomial preconditioning and deflation were not used. 
\begin{table}
\caption{ Trace results for $24^4$ lattices with error tolerance $0.0005*24^4$. Compare Hutchinson with the Multipolynomial Method. We average with 10 configurations.}
\begin{center}
\begin{tabular}{|c|c|c|c|c|c|}  \hline\hline
Method                              & Noise vectors     & Time          & MVP's             \\ \hline \hline
Hutch., defl. PP(70)-GMRES          & 188               & 19.6 hours    & $2.02*10^5$       \\ \hline
3 Poly's, $10^{-12}$ for deflation  & 2, 17.1, 101      & 6.84 hours    & $6.68*10^4$       \\ \hline
3 Poly's, extra deflation           & 2, 3, 24.9        & 3.07 hours    & $2.80*10^4$       \\ \hline \hline
\end{tabular}
\end{center}
\label{Tab:Tr3}
\end{table}

\section{Conclusions}

We have proposed a new approach to computing the trace of the inverse of a large QCD matrix. It consists of a multilevel Monte Carlo method with different degree polynomials.  The polynomials give approximations of different accuracy to the inverse of the matrix. The lower degree polynomials are from GMRES and the highest degree polynomial is from polynomial preconditioned GMRES. 

This is the first use of polynomials for multilevel Monte Carlo with nonsymmetric matrices. For this method to be effective, it is essential to deflate eigenvalues from the Monte Carlo steps that have a difference of two polynomials. This deflation is also new, as is the use of a deflated polynomial to lower the degree of the highest degree polynomial. Also important to the method is that this highest degree polynomial is a double polynomial with much less orthogonalization expense needed to generate it. Also implemented is a deflated version of polynomial preconditioned GMRES that is very efficient and is new to QCD calculations. Computation of eigenvalues and eigenvectors is done with a polynomial preconditioned Arnoldi method that makes deflation much more practical. Putting the pieces together gives a multipolynomial Monte Carlo method that is very efficient.

We plan future work on this method, including automatic optimization of the cost function~\cite{Gi15} as well as using nonsymmetric Lanczos to develop the polynomials and to compute eigenvalues. This will likely give a lower degree polynomial than the current double polynomial. For QCD matrices, this can be implemented efficiently; see Ref.~\cite{FroNoeetal95}.

\section{Acknowledgements}

We acknowledge the use of the Baylor High Performance Cluster as well as the Texas Advanced Computing Center (TACC). We also acknowledge the Baylor Summer Research Award program, the previous work of Suman Baral, and the QQCD code from Randy Lewis.

\end{document}